\documentclass{elsart}
\usepackage{epsfig,amssymb,amsmath}
\usepackage{graphicx}

\journal{Chaos, Solitons \& Fractals}

\begin{document}

\begin{frontmatter}

\title{Travelling wave solutions of the generalized Benjamin-Bona-Mahony equation}

\author{P. G. Est\'evez}
\address{Departamento de F\'{\i}sica Fundamental, \'Area de F\'{\i}sica
Te\'{o}rica, Universidad de Salamanca, 37008 Salamanca, Spain}

\author{\c{S}. Kuru\corauthref{abs1}},
\corauth[abs1]{\emph{On leave of absence from:} Department of
Physics, Faculty of Science, Ankara University 06100 Ankara,
Turkey.}
\author{J. Negro} and
\ead{jnegro@fta.uva.es}
\author{L. M. Nieto}
\address{Departamento de F\'{\i}sica Te\'orica, At\'omica y \'Optica,
Universidad de Valladolid, 47071 Valladolid, Spain}

\begin{abstract}
A class of particular travelling wave solutions of the generalized
Benjamin-Bona-Mahony equation is studied systematically using the
factorization technique. Then, the general travelling wave solutions
of Benjamin-Bona-Mahony equation, and of its modified version, are
also recovered.\\
Pacs:{05.45.Yv,  52.35.Mw,  52.35.Sb, 02.30.Jr}
\end{abstract}

\end{frontmatter}

\section{Introduction}
The generalized BBM (Benjamin-Bona-Mahony) equation has a higher order
nonlinearity of the form
\begin{equation}\label{gb}
u_{t}+u_{x}+ a\,u^{n}\,u_{x}+ u_{xxt}=0, \quad n\geq1,
\end{equation}
where $a$ is constant. The case $n=1$ corresponds to the BBM
equation
\begin{equation}\label{b}
u_{t}+u_{x}+ a\,u\,u_{x}+ u_{xxt}=0,
\end{equation}
which was first proposed in 1972 by Benjamin et al \cite{benjamin}.
This equation is an alternative to the Korteweg-de Vries  (KdV)
equation, and describes the unidirectional propagation of
small-amplitude long waves on the surface of water in a channel. The
BBM equation is not only convenient  for shallow water waves but
also for hydromagnetic waves, acoustic waves, and therefore it has
more advantages compared with the KdV equation. When $n=2$,
Eq.(\ref{gb}) is called the modified BBM equation
\begin{equation}\label{mb}
u_{t}+u_{x}+ a\,u^{2}\,u_{x}+ u_{xxt}=0.
\end{equation}
When looking for travelling wave solutions, the BBM and modified BBM
equations can be reduced to ordinary differential equations that
possess the Painlev\'{e} property and which are integrable in terms
of elliptic functions \cite{clarkson,ince}. The generalized BBM
equation is also integrable in terms of elliptic functions, provided
that some restrictions on the parameters are imposed. Recently many
methods have been presented to obtain the travelling wave solutions
of the generalized BBM equation: the tanh-sech and the sine-cosine
methods \cite{wazwaz,wazwaz1}, an approach based on balancing
principle to obtain some explicit solutions in terms of elliptic
function \cite{nickel}, and an extended algebraic method with
symbolic computation \cite{tang}.

Our aim is to investigate systematically the travelling wave
solutions of these equations, applying the factorization technique
\cite{pilar,Perez}. Thus, we will get all the previously known
solutions and some new ones, supplying a new general approach.
Assuming that the generalized BBM equation has an exact solution in
the form of a travelling wave, then it will reduce to a third order
ordinary differential equation (ODE). This equation can be
integrated trivially to a second order ODE, which can be factorized
in two ways: the first one by means of differential operators and
the second one by using a first integral (that can also be
factorized in terms of first integrals). These factorizations give
rise to the same first order ODE that provides the travelling wave
solutions of the nonlinear equation. This first order ODE for $n=1$
and $n=2$ is integrable, but for other values of $n$, we can also
find some particular solutions by imposing some restrictions on the
parameters.

The paper is organized as follows. In section 2, we introduce the
factorization technique for nonlinear equations, and we show how to
apply it to find the travelling wave solutions of the generalized
BBM equation. In section 3, we consider some special cases to get
particular solutions of generalized BBM. In section 4, we obtain the
solutions of the BBM and modified BBM. Finally, section 5 ends the
paper with some conclusions. In the Appendix, we give also some
useful information about the elliptic functions that are used in the
previous sections.

\section{Travelling waves of the generalized Benjamin-Bona-Mahony equation}
\subsection{Travelling wave solutions}
Let us assume that Eq. (\ref{gb}) has an exact solution in the form
of a travelling wave
\begin{equation}\label{xi}
u(x,t)=\phi(\xi),\quad \xi=h\,x-\omega\,t,
\end{equation}
where $h$ and $\omega$ are real constants. If we substitute
(\ref{xi}) in Eq. (\ref{gb}), we get
\begin{equation}\label{or}
-h^2\,\omega\,\phi_{\xi\xi\xi}+ (h-\omega) \phi_{\xi}+
h\,a\,\phi^{n}\,\phi_{\xi}=0.
\end{equation}
After integrating with respect to $\xi$, we have
\begin{equation}\label{or1}
\phi_{\xi\xi}- \frac{ h-\omega}{h^{2}\,\omega}\,\phi- \frac{
a}{(n+1)\,h\,\omega}\,\phi^{n+1}=-R,
\end{equation}
where $R$ is an integration constant. Let us introduce the following linear
transformation of the dependent and independent variables
\begin{equation}\label{lt}
 \xi=h\,\theta, \quad \phi(\xi)= \left(\frac{c\,(n+1)}{a}\right)^{1/n} W(\theta)
\end{equation}
where $\theta=x-c\,t$ and $c={\omega}/{h}$. In this way (\ref{or1})
becomes the nonlinear second order ODE
\begin{equation}\label{or2}
 \frac{d^2 W}{d \theta^2}-W^{n+1}-k\,W=D,
\end{equation}
where the new constants are
\begin{equation}\label{kd}
k=\frac{1-c}{c}, \quad D= {-R\,h^2}
\left(\frac{a}{c\,(n+1)}\right)^{1/n}.
\end{equation}
Therefore, if we are interested in finding the travelling wave solutions of (\ref{gb}),
we have to solve the ODE (\ref{or2}).

\subsection{Factorization of some special type of  nonlinear second order ODE}
In this section we will introduce a factorization technique applied
to nonlinear second order ODE of the special form
\begin{equation}\label{9}
 \frac{d^2 W}{d \theta^2}-\beta \frac{d W}{d \theta}+F(W)=0,
\end{equation}
where $F(W)$ is an arbitrary function of $W$ and $\beta$ is
constant. This equation can be factorized as
\begin{equation}\label{10}
\left[\frac{d}{d \theta}-f_2(W,\theta)\right]\left[\frac{d}{d
\theta}-f_1(W,\theta)\right] W(\theta)=0
\end{equation}
being $f_1$ and $f_2$ two unknown functions that may depend explicitly on $W$ and $\theta$.
In order to find $f_1$ and $f_2$, we expand (\ref{10})
\begin{eqnarray}
\frac{d^2 W}{d \theta^2}&-&\left(f_1 +f_2 +\frac{\partial
f_1}{\partial W} \ W \right)\frac{d W}{d \theta}+f_1 \,f_2 \,W - W\
\frac{\partial f_1}{\partial \theta}=0, \label{11}
\end{eqnarray}
and then comparing with (\ref{9}), we obtain the following
consistency conditions
\begin{equation}\label{12}
f_1f_2=\frac{F}{W}+\frac{\partial f_1}{\partial \theta},
\end{equation}
\begin{equation}\label{13}
f_2+\frac{\partial(W f_1)}{\partial W}=\beta.
\end{equation}

If we find a solution for this factorization problem, it will allow
us to write a compatible first order ODE
\begin{equation}\label{14}
\left[\frac{d}{d \theta}-f_1(W,\theta)\right] W(\theta)=0
\end{equation}
that provides a (particular) solution \cite{pilar,Perez} to the nonlinear ODE (\ref{9}).
In the applications of this
paper $f_{1}$ and $f_{2}$ will depend only on $W$.

\subsection{Factorization of the generalized BBM equation}\label{2.3}
When we apply the factorization technique described above  to Eq. (\ref{or2}),
then  we have $\beta=0$,
$F(W)=-W^{n+1}-k\,W-D$,  and the consistency conditions
given by (\ref{12}) and (\ref{13}) take the form
\begin{equation}\label{c1}
f_1f_2=-W^n-k -D \,W^{-1},
\end{equation}
\begin{equation}\label{c2}
f_2=-f_1 -W\frac{\partial f_1}{\partial W}.
\end{equation}
Substituting (\ref{c2}) in (\ref{c1}) we get
\begin{equation}\label{eqf}
W f_1 \frac{\partial f_1}{\partial W}+f_{1}^2-W^n-k-D\, W^{-1}=0.
\end{equation}
The solutions of this equation are
\begin{equation}\label{f}
f_1=\pm\sqrt{\frac{2\,W^n}{n+2}+k+\frac{2\,D}{W}+\frac{C}{W^2}},
\end{equation}
where $C$ is an integration constant. By replacing $f_1$ given by
(\ref{f}) in Eq. (\ref{14}), we have the first order ODE
\begin{equation}\label{fu}
\frac{d W}{d \theta}\mp
\sqrt{\frac{2\,W^{n+2}}{n+2}+k\,W^2+2\,D\,W+C}=0.
\end{equation}
As we have already mentioned, when we solve this equation, we get
also a particular solution of (\ref{or2}).

In the special case of the generalized BBM equation, another way to
get this result is as follows: if Eq. (\ref{or2}) is
multiplied by $2\,W'$ and  integrated, we arrive to
\begin{eqnarray*}\label{fact}
&& \left[\frac{d W}{d \theta} -
\sqrt{\frac{2\,W^{n+2}}{n+2}+k\,W^2+2\,D\,W+C_{0}}\right] \\&&\times
\left[ \frac{d W}{d \theta} +
\sqrt{\frac{2\,W^{n+2}}{n+2}+k\,W^2+2\,D\,W+C_{0}}\right]= 0
\end{eqnarray*}
which is equivalent to (\ref{fu}). Remark that the last equation is
a product of functions, while the factorization (\ref{10}) is a
product of operators.

These factorizations are valid for any value of the parameters
($k,n$) and the integration constants ($C, D$). However, the
integrability of this equation is obtained, in general, only if some
constraints are imposed.
As a result, it can be said that the factorization does not imply
the integrability of a nonlinear equation, but can produce some
solutions, under certain parameter restrictions (to be determined
later, when solving Eq. (\ref{fu})).

If we want to guarantee the integrability of (\ref{fu}), the powers
of $W$ have to be integer numbers between $0$ and $4$, and therefore
$n\in\{-1,0,1,2\}$ \cite{ince}. [Nevertheless, recall that an
initial assumption on the generalized BBM equation \eqref{gb} is
$n\geq 1$.] On the other side, equation (\ref{or2}) has the
Painlev\'{e} property only if the same condition on $n$ is satisfied
\cite{ince}. In these cases Eq. (\ref{or2}) is integrable in terms
of elliptic functions.

In order to find other values of $n$ for which it is possible to find particular solutions,
let us make in Eq. (\ref{fu}) the transformation $W=\varphi^p, p\neq 0,1$, getting
\begin{eqnarray}
\left(\frac{d\varphi}{d \theta}\right)^2 &=&
\frac{2}{(n+2)\,p^2}\varphi^{2+n\,p} +\frac{k}{p^2}\,\varphi^{2}+
\frac{2\,D}{p^2}\,\varphi^{2-p}+\frac{C}{p^2}\,\varphi^{2-2p}.
\label{phi}
\end{eqnarray}
This equation is of the same type as the initial one  (\ref{fu}),
hence the same integrability conditions are valid. In this way we
have the following additional cases, that will require to impose
some conditions on the parameters (with $n\geq 1$):
\begin{enumerate}
\item
If $C=D=0$, then
$$
p\in\{-\frac2n, -\frac1n, \frac1n,\frac2n\}.
$$
In the next section it will be proved that $p=1/n$ and $p=2/n$ give the same solution
$u(x,t)$ of the BBM equation \eqref{gb}, and therefore we consider only the case $p=1/n$.
Then, Eq. (\ref{phi}) takes the form
\begin{equation}\label{phia}
\left(\frac{d\varphi}{d \theta} \right)^2=\frac{2\,n^2}{(n+2)}\varphi^{3}+k\,
n^2\,\varphi^{2}.
\end{equation}
The same discussion applies for the values $p=-1/n$, $p=-2/n$, and
choosing $p=-1/n$, Eq. (\ref{phi}) becomes
\begin{equation}\label{phib}
\left(\frac{d\varphi}{d \theta}
\right)^2=\frac{2\,n^2}{(n+2)}\varphi+k\, n^2\varphi^{2}.
\end{equation}
\item
If $C\neq0$ and $D=0$, we have $n=4$ and two possibilities: either $p=1/2$ with
\begin{equation}\label{phic}
\left(\frac{d\varphi}{d\theta} \right)^2=\frac{4}{3}\,\varphi^{4}+4\,k\,\varphi^{2}+4\,C\,\varphi,
\end{equation}
or $p=- {1}/{2}$ with
\begin{equation}\label{phiccc}
\left(\frac{d\varphi}{d\theta}
\right)^2=\frac{4}{3}+4\,k\,\varphi^{2}+4\,C\,\varphi^3.
\end{equation}
\item
If $C=0$ and $D\neq 0$, no new solutions appear.
\end{enumerate}

The solutions of equations (\ref{phia})--(\ref{phiccc}) can be
expressed in terms of elliptic functions. Then, using (\ref{xi}),
(\ref{lt}), (\ref{kd}),  and recalling
$W(\theta)=\varphi^p(\theta)$, the particular solution of the
generalized BBM equation (\ref{gb}) reads
\begin{equation}\label{sgb}
u(x,t)=\left(\frac{c\,(n+1)}{a}\right)^{1/n}\varphi^{p}(x-c\,t).
\end{equation}
The details are given in the next section.

\section{Solutions of the generalized Benjamin-Bona-Mahony equation}
In this section, using the basic results on elliptic functions shown in the Appendix,
we will analyze the solutions of the four equations (\ref{phia})--(\ref{phiccc})
obtained before, which produce particular solutions of the BBM equation.

\subsection{Case $C=D=0$, $p={1}/{n}$}\label{3.1}
In this case the third order polynomial of  Eq.~(\ref{phia})
\begin{equation}\label{f1}
f(\varphi)=\frac{2\,n^2}{(n+2)}\,\varphi^{3}+k\,n^2\,\varphi^{2}
\end{equation}
has two roots: $\varphi_{0}=0$ (with multiplicity two) and
$\varphi_{0}=-{k\,(n+2)}/{2}$. When we substitute the derivatives of
$f(\varphi)$ in terms of $\varphi_0$ in Eq.~(\ref{x}), we get the
following solution for Eq.~(\ref{phia})
\begin{equation}\label{gs1}
\varphi=\frac{12 \,\varphi_{0}\,\wp(\theta;g_2,g_3)+ \frac{12\,n^2}{n+2}\,
\varphi_{0}^{2}+5\,k\,n^2 \varphi_{0}}{ 12 \,\wp(\theta;g_2,g_3) - \frac{6
\,n^2}{n+2}\,\varphi_{0} -k\,n^{2}},
\end{equation}
where the invariants are
\begin{equation}\label{g1}
g_{2}=\frac{k^2\,n^4}{12}, \quad\quad  g_{3}=-\frac{k^3\,n^6}{216}.
\end{equation}
The root $\varphi_0=0$ gives the trivial solution, $\varphi=0$, and the nonzero solution of
Eq.~(\ref{phia}) can be easily found replacing $\varphi_{0}=-{k(n+2)}/{2}$ in (\ref{gs1}):
\begin{equation}\label{s1}
\varphi=\frac{k(n+2)}{4}\left(\frac{kn^2-12\,\wp(\theta;g_2,g_3)}{kn^2+
6\,\wp(\theta;g_2,g_3)}\right)
\end{equation}
or using relation (\ref{hp}) in (\ref{s1})
\begin{equation}\label{s11}
\varphi=\frac{2(n+2)}{n^2}\wp(\theta+\omega;g_2,g_3)-\frac{k\,(n+2)}{6}
\end{equation}
being $\omega$ the half-period. For the values $g_{2}$ and $g_{3}$ of
(\ref{g1}), the discriminant defined by (\ref{di}) is equal to zero.
Hence, we have both solitary wave and periodic solutions. When we
use (\ref{sh}) in (\ref{s1}), we have the solitary wave solution
\begin{equation}\label{ss1}
\varphi=-\frac{k\,(n+2)}{2}\ {\rm sech}^{2}
\left[\frac{n}{2}\sqrt{k}\,\theta\right]
\end{equation}
for $0<c<1$. We get the same result by making use of (\ref{s11})
with the half-period $\omega'=(i\pi/\sqrt{k}n)$. Using half-angle
formulas for $\cosh x$, solution (\ref{ss1}) can also be written
as
\begin{equation}\label{ss2}
\varphi=-k\,(n+2)\frac{1}{1+ \cosh  [n\sqrt{k}\,\theta ]}.
\end{equation}
When we use (\ref{sn}) in (\ref{s1}), we have the periodic solution
\begin{equation}\label{ss3}
\varphi=\frac{k\,(n+2)}{2} \ \sec^{2}
\left[\frac{n}{2}\sqrt{-k}\,\theta\right]
\end{equation}
for $c>1$. The same result can be obtained  from (\ref{s11}) with
the half-period $\omega=(\pi/\sqrt{-k}n)$. This solution can also be
rewritten in the form
\begin{equation}\label{ss4}
\varphi=k\,(n+2)\frac{1}{1+\cos[n\,\sqrt{-k}\,\theta]}.
\end{equation}
Finally,  substituting (\ref{ss1})--(\ref{ss4}) in (\ref{sgb}) and
taking into account (\ref{kd}), we get the following solutions of
Eq.~(\ref{gb})
\begin{equation}\label{sgbs}
\begin{array}{ll}
u(x,t)&= \displaystyle \left(\frac{(n+1)(n+2)(c-1)/a}{1+\cosh\left[n
\sqrt{\frac{1-c}{c}}(x-c\,t)\right]}\right)^{\!\!1/n}
\\[5.ex]
&= \displaystyle \left(\frac{(n+1)(n+2)(c-1)}{2a} \right)^{\!\!1/n}
\left({\rm sech}^{2}\left[\frac{n}{2}
\sqrt{\frac{1-c}{c}}(x-c\,t)\right]\right)^{\!\!1/n} \end{array}
\end{equation}
(solitary waves) for $c<1$, and
\begin{equation}\label{sgbp}
\begin{array}{ll}
u(x,t)&= \displaystyle \left(\frac{(n+1)(n+2)(1-c)/a}{1+\cos\left[n
\sqrt{\frac{c-1}{c}}(x-c\,t)\right]}\right)^{\!\!1/n}
\\[5.ex]
&=\displaystyle \left(\frac{(n+1)(n+2)(1-c)}{2a}\right)^{\!\!1/n}
\left(\sec^{2}\left[\frac{n}{2}
\sqrt{\frac{c-1}{c}}(x-c\,t)\right]\right)^{\!\!1/n} \end{array}
\end{equation}
(periodic) for $c>1$.

We will show now that, as it was mentioned before, by choosing in Eq.~\eqref{phi} $p={2}/{n}$
instead of $p={1}/{n}$,
we will recover the same solutions for (\ref{gb}). In this case,
Eq.~\eqref{phi} becomes
\begin{equation}\label{f2}
\left(\frac{d\varphi}{d\theta} \right)^{2}=\frac{n^2}{2\,(n+2)}\
\varphi^{4}+\frac{k\, n^2}{4}\ \varphi^{2}
\end{equation}
and the forth order polynomial has three different roots: $\varphi_{0}=0$ (with
multiplicity two) and
\begin{equation}\label{pi0b}
\varphi_{0}=\pm\sqrt{\frac{-k\,(n+2)}{2}}.
\end{equation}
From (\ref{x}), the solution of Eq.~(\ref{f2}) for each root
($\varphi_0$) can be found
\begin{equation}\label{gs2}
\varphi=\frac{48\, \varphi_{0}\,\wp(\theta;g_2,g_3) + \frac{12\,n^2}{n+2}\,
\varphi_{0}^{3}+5\,k\,n^2 \,\varphi_{0}}{ 48\,\wp(\theta;g_2,g_3) -k\,n^{2}
- \frac{12\,n^2}{n+2}\,\varphi_{0}^{2}}
\end{equation}
where the invariants are given by
\begin{equation}\label{g2}
g_{2}=\frac{k^2\,n^4}{192}, \quad\quad
g_{3}=-\frac{k^3\,n^6}{13824}.
\end{equation}
When we choose $\varphi_0=0$, we have again the trivial solution
$\varphi=0$, but if we use the roots given in Eq.~(\ref{pi0b}), we
get the solutions of Eq.~(\ref{f2})
\begin{equation}\label{s2}
\varphi^{\pm}=\mp\sqrt{\frac{-k\,(n+2)}{2}}\frac{k\,n^2-48\,\wp(\theta;g_2,g_3)}
{5\,k\,n^2+48\,\wp(\theta;g_2,g_3)}.
\end{equation}
For $g_2$ and $g_3$ given in \eqref{g2}, the discriminant is equal
to zero, $\Delta=0$, so the Weierstrass function degenerates into
trigonometric and hyperbolic functions. Then, taking into account
(\ref{sh}) and (\ref{sn}), it is easy to see that the solutions
(\ref{s2}) give rise to the same solitary wave and periodic
solutions of Eq. (\ref{gb}) as (\ref{sgbs}) and (\ref{sgbp}).
\subsection{Case $C=D=0$, $p=-{1}/{n}$}
Here we have the second order polynomial
\begin{equation}\label{phibp}
f(\varphi)=\frac{2\,n^2}{(n+2)}\, \varphi+k\, n^2\,\varphi^{2}
\end{equation}
with two roots: $\varphi_{0}=0$ and
\begin{equation}\label{mphi0}
\varphi_{0}=-\frac{2}{k\,(n+2)}.
\end{equation}
The expression for the solutions of (\ref{phib}) in terms of
$\varphi_{0}$ is
\begin{equation}\label{mgs}
\varphi=\frac{12\, \varphi_{0}\,\wp(\theta;g_2,g_3)+ \frac{6\,n^2}{n+2}+5
k\,n^2 \, \varphi_{0}}{ 12 \wp(\theta;g_2,g_3) - k\,n^{2}}
\end{equation}
where the invariants are given by (\ref{g1}).
Taking the root $\varphi_{0}=0$ in  (\ref{mgs}), we get
\begin{equation}\label{mgs1}
\varphi=\frac{6\,n^{2}}{12(n+2)\,\wp(\theta;g_2,g_3)-k\,n^{2}(n+2)}.
\end{equation}
Since the discriminant is equal to zero  for $g_{2}$ and $g_{3}$
given by (\ref{g1}), we can express the Weierstrass function in
terms of trigonometric and hyperbolic functions. Thus, substituting
(\ref{sh}) in (\ref{mgs1}), the solitary wave solution can be
written
\begin{equation}\label{ms1}
\varphi=-\frac{2}{k\,(n+2)}\ {\rm
sinh}^{2}\left[\frac{n}{2}\sqrt{k}\,\theta\right]
\end{equation}
for $c<1$, and using (\ref{sn}) in (\ref{mgs1}), we have the
periodic solution
\begin{equation}\label{mms1}
\varphi=\frac{2}{k\,(n+2)}\ \sin^{2}
\left[\frac{n}{2}\sqrt{-k}\,\theta\right]
\end{equation}
for $c>1$.

When we take the second root (\ref{mphi0}) in (\ref{mgs}), we have
\begin{equation}\label{mgs2}\nonumber
\varphi=\frac{4\,k\,n^{2}+ 24\,\wp(\theta;g_2,g_3)}{
k\,(n+2)(k\,n^{2}-12\,(n+2)\,\wp(\theta;g_2,g_3))}.
\end{equation}
Having in mind the degenerate cases of the Weierstrass function
(shown in the Appendix), this solution can be expressed as
\begin{equation}\label{ms2}
\varphi=-\frac{2}{k\,(n+2)}\ \cosh^{2}
\left[\frac{n}{2}\sqrt{k}\,\theta\right]
\end{equation}
for $c<1$ and
\begin{equation}\label{mms2}
\varphi=\frac{2}{k\,(n+2)}\ \cos^{2}
\left[\frac{n}{2}\sqrt{-k}\,\theta \right]
\end{equation}
for $c>1$. Substituting (\ref{ms1})--(\ref{mms2}) in (\ref{sgb}) and
taking into account (\ref{kd}), we get the solutions of
Eq.~(\ref{gb}) which were given by (\ref{sgbs}) and (\ref{sgbp}). In
addition, it can be also proved that the choice $p=- {2}/{n}$
 in Eq.~\eqref{phi} gives exactly the same solutions for Eq.~(\ref{gb}).

\subsection{Case $C\neq0$, $D=0$, $n=4,\,p={1}/{2}$}
Now the quartic polynomial
\begin{equation}\label{f3}
f(\varphi)=\frac{4}{3}\,\varphi^{4}+4\,k\,\varphi^{2}+4\,C\,\varphi
\end{equation}
has four roots. Substituting the first and second derivative of
(\ref{f3}) in terms of $\varphi_0$ in (\ref{x}), we have the general
expression for the solutions
\begin{equation}\label{gs3}
\varphi=\frac{3\,\varphi_{0}\,\wp(\theta;g_2,g_3)+ 2\,\varphi_{0}^3+
5\,k\,\varphi_{0}+3\,C}
{3\,\wp(\theta;g_2,g_3)-2\varphi_{0}^2-k}
\end{equation}
where the invariants are given by
\begin{equation}\label{gs3}
g_{2}=\frac{4\,k^2}{3},\quad\quad g_{3}=-\frac{8\,k^3}{27}-\frac{4
C^2}{3}.
\end{equation}
When we take the simplest root of $f(\varphi)$, $\varphi_{0}=0$, the
solution of Eq.~(\ref{phic}) is
\begin{equation}\label{s3}
\varphi=\frac{C}{\wp(\theta;g_2,g_3)-{k}/{3}}.
\end{equation}
Replacing (\ref{s3}) and (\ref{kd}) in (\ref{sgb}), we get the
particular solution of Eq.~(\ref{gb})
\begin{equation}\label{sgb3}
u(x,t)=\left(\sqrt{\frac{5\,c}{a}}\frac{3 \,c \,C}{3\,
c\,\wp(x-c\,t)+c-1}\right)^{1/2}.
\end{equation}
Using the other roots of (\ref{f3}) in Eq.~(\ref{x}), we can get the
other solutions, that become trigonometric and hyperbolic, like the
solutions of the case \ref{3.1} for $C=0$.

\subsection{Case $C\neq0$, $D=0$, $n=4,\,p=-{1}/{2}$}
In this case
we have the third order  polynomial
\begin{equation}\label{mf3}
f(\varphi)=4\,C\,\varphi^3+4\,k\,\varphi^{2}+\frac{4}{3}
\end{equation}
has three roots. Here we will not give the roots of this polynomial,
since they are a bit cumbersome. But, substituting the first and
second derivative of (\ref{mf3}) in terms of $\varphi_0$ in
(\ref{x}), we have the general expression for the solutions
\begin{equation}\label{mgs3}
\varphi=\frac{3\,\varphi_{0}\,\wp(\theta;g_2,g_3)+5\,k\,\varphi_{0}+6\,C\,\varphi_{0}^2}
{3\,\wp(\theta;g_2,g_3)+3\,C\,\varphi_{0}-k}
\end{equation}
where the invariants given by (\ref{gs3}) and the discriminant
$\Delta\neq0$.

We also notice that the choice $n=4,\,p=-{1}/{2}$ gives rise to the
same particular solution (\ref{sgb3}) for Eq.~(\ref{gb}) and the
other solutions can be found by the same procedure as in the above
cases.

\section{Solutions of the BBM and modified BBM}
In this section we will consider the solutions obtained for $n=1, 2$
when the integration constant $D$ of Eq.~(\ref{or2}) is such that $D\neq0$
(to avoid confusion, in this section we will use the notation $D\equiv D_n$,
because, as we will immediately see, this constant is chosen to be dependent on $n$).

First of all, let us transform Eq.~(\ref{or2}) through the change of
function $W(\theta)=U(\theta)+\delta$ into
\begin{eqnarray}\label{ord}
&&\frac{d^2 U}{d \theta^2}-\left(U^{n+1}+\frac{(n+1)!}{n!}U^n
\delta+ \cdots \right. \\
&&\quad \left. + \frac{(n+1)!}{2(n-1)!}U^{n-1}
\delta^2+...+\frac{(n+1)!}{n!}U\delta^{n}\right)= k\,U \nonumber 
\end{eqnarray}
where the integration constant $D_n$ has been chosen as
\begin{equation}\label{d}
 D_n=-k \,\delta-\delta^{n+1}.
\end{equation}
Remark that this change of variable does not give any restriction on
the solution of (\ref{or2}).

\subsection{BBM equation ($n=1$)} We can find the solutions of Eq.~(\ref{b})
when the integration constant is non-zero, $D\neq0$, in
Eq.~(\ref{or2}). To do this, first we take $n=1$ and $\delta\neq0$
in Eq.~(\ref{ord})
\begin{equation}\label{ordn1}
\frac{d^2 U}{d \theta^2}-U^{2}+(2\,\delta-k)U=0
\end{equation}
with
\begin{equation}\label{d1}
 D_1=-k\,\delta-\delta^{2}.
\end{equation}
For this case, the third order polynomial is
\begin{equation}\label{f11}
f(U)=\frac{2}{3}U^{3}-(2\,\delta-k)U^{2}+C_{1}
\end{equation}
with three different roots. When we follow the same procedure as
mentioned above, we get the solutions of the second order ODE
(\ref{ordn1}) for all roots
\begin{equation}\label{gs11}  \nonumber
U(\theta)=-\frac{5 (k -2\,\delta)U_{0}+4\,U_{0}^2+ 12\,U_{0}\,
\wp(\theta;g_2,g_3)}{(k-2\,\delta)+2\,U_{0}-12\,\wp(\theta;g_2,g_3)},
\end{equation}
where
\begin{equation}\label{gb1}
g_{2}=\frac{(k-2\,\delta)^2}{12},\quad
g_{3}=-\frac{(k-2\,\delta)^3}{216}-\frac{C_1}{36},
\end{equation}
and $\Delta\neq0$. Then the solutions of Eq.~(\ref{b}) can be found
from the relation
\begin{equation}\label{sgb1}
u(x,t)=\frac{2\,c}{a}(U(\theta)+\delta).
\end{equation}
When we choose $C_1=0$, the polynomial (\ref{f11}) has two different
roots: $0$ (with multiplicity two) and $U_0=-3(k-2\,\delta)/2$. For
the nonzero root we have the following solution
\begin{equation}\label{sb1}
U(\theta)=6\,\wp(\theta+\omega;g_2,g_3)-\frac{(k-2\,\delta)}{2}
\end{equation}
where
\begin{equation}\label{gb11}
g_{2}=\frac{(k-2\,\delta)^2}{12},\quad
g_{3}=-\frac{(k-2\,\delta)^3}{216}
\end{equation}
with $\Delta=0$. Therefore, we can express the Weierstrass function as
hyperbolic (\ref{sh}) and trigonometric (\ref{sn}) forms. Then,
substituting $U(\theta)$ given by (\ref{sb1}) in (\ref{sgb1}),
having in mind the simplified form of the Weierstrass function, and
choosing $\delta=3\,k/4$ where $k=(1-c)/c$,
 $\theta=x-c\,t$, we have solitary wave (dark soliton) solutions for $c>1$
\begin{equation}\label{sgb1k} \nonumber
u(x,t)= \frac{3(c-1)}{2a}\ \tanh^{2}\left[\frac{1}{2}
\sqrt{\frac{c-1}{2c}}(x-c\,t)\right]
\end{equation}
and periodic singular solutions for $c<1$
\begin{equation}\label{sgb1p} \nonumber
u(x,t)= \frac{3(1-c)}{2a}\ \tan^{2} \left[\frac{1}{2}
\sqrt{\frac{1-c}{2c}}(x-c\,t) \right].
\end{equation}

Here we can also consider the special value  $\delta=0$. In this
case we have
\begin{equation}\label{f5}
\left(\frac{dU}{d\theta} \right)^2=\frac{2}{3}U^{3}+k\,U^{2}+C_{1}
\end{equation}
and the third order polynomial has three different roots.
Taking into account  Eq.~(\ref{x}), the solution of (\ref{f5}) for
each root takes the form
\begin{equation}\label{gs5}
U=-\frac{5\,k\,U_{0}+4\,U_{0}^2+ 12\,U_{0}
\wp(\theta;g_2,g_3)}{k+2\,U_{0}-12\,\wp(\theta;g_2,g_3)}
\end{equation}
where
\begin{equation}\label{g5}
g_{2}=\frac{k^2}{12},\quad\quad
g_{3}=-\frac{k^3}{216}-\frac{C_{1}}{36}.
\end{equation}
If  $C_{1}\neq0$, the discriminant of $g_2$ and $g_3$ is different
from zero, and therefore the solutions of Eq.~(\ref{gs5}) can not be
simplified. Since we are interested in solitary wave and periodic
solutions, we can choose $C_{1}=0$. Then, the invariants take the
form of (\ref{g1})
with $n=1$ and $\Delta=0$. Thus, the third order polynomial has two
different roots: $0$ (twice) and $U_{0}=-3k/2$. While the root
$U_{0}=0$ gives the trivial solution of Eq.~(\ref{f5}), $U=0$, the
nonzero root implies the following solution
\begin{equation}\label{s5}
U=\frac{3k}{4}\ \frac{2 k -12\,\wp(\theta;g_2,g_3)}{k+6\,\wp(\theta;g_2,g_3)}
\end{equation}
or from (\ref{hp})
\begin{equation}\label{s55}
U=6 \wp(\theta+\omega;g_2,g_3)-\frac{k}{2}
\end{equation}
where $\omega$ is a half period. Now it is easy to check that
(\ref{s5}) and (\ref{s55}) correspond to (\ref{s1}) and (\ref{s11})
with $n=1$, respectively. Therefore, the particular solutions of
Eq.~(\ref{b}) can be found substituting $n=1$ in (\ref{sgbs}) and
(\ref{sgbp}).

\subsection{Modified BBM equation ($n=2$)}
Equation (\ref{gb}) with $n=2$ reduces to the modified BBM equation
(\ref{mb}). Then, Eq.~(\ref{ord}) becomes
\begin{equation}\label{ordn2}
\frac{d^2 U}{d \theta^2}-(U^{3}+3\,\delta\,U^{2}+(\delta^{2}+k)U)=0
\end{equation}
with
\begin{equation}\label{d2}
 D_2=-k\,\delta-\delta^{3}.
\end{equation}
To obtain the solution of (\ref{ordn2}), we have to solve following
first order equation that can be seen from the Section 2.3
\begin{equation}\label{ford}
\frac{d U}{d
\theta}\mp\sqrt{\frac{1}{2}\,U^{4}+2\,\delta\,\,U^{3}+K\,
U^{2}+C_{2}}=0
\end{equation}
where $K=3\,\delta^{2}+k$. For this case we have the forth order
polynomial
\begin{equation}\label{f22}
f(U)=\frac{1}{2}U^{4}+2\,\delta U^{3}+K\,U^{2}+C_{2}
\end{equation}
with four different roots. The solution of Eq.~(\ref{ford}) for each
root can be obtained, applying the same procedure mentioned above,
as
\begin{equation}\label{mbs}
U(\theta)=\frac{5\,K\, U_{0}+3\,U_{0}^3+ 12\,\delta
\,U_{0}^{2}+12\,U_{0}\,\wp(\theta;g_2,g_3)}{12\,\wp(\theta;g_2,g_3)-
k-3\,(\delta+U_{0})^2}
\end{equation}
where
\begin{equation}\label{gm1}
g_{2}=\frac{K^2}{12}+\frac{C_2}{2},\qquad
g_{3}=-\frac{K^3}{216}+\frac{C_2\,K}{12}-\frac{C_{2}\,\delta^2}{4},
\end{equation}
and $\Delta\neq0$. Now, the solution of Eq.~(\ref{mb}) can be
obtained from
\begin{equation}\label{sgb2}
u(x,t)=\sqrt{\frac{3\,c}{a}}\,(U (\theta)+ \delta).
\end{equation}

If $C_2=0$ and $\delta\neq0$, the forth order polynomial (\ref{f22})
has three different roots: $0$ (with multiplicity two) and
\begin{equation}\label{u0}
U_{0}^{\pm}=-2\,\delta\pm\sqrt{-2\,(k+\delta^2)}.
\end{equation}
In this case $\Delta=0$ and the invariants are
\begin{equation}\label{gm11}
g_{2}=\frac{K^2}{12},\qquad g_{3}=-\frac{K^3}{216}.
\end{equation}
When we substitute these roots and the degenerate forms of the
Weierstrass function in Eq.~(\ref{mbs}), we have
\begin{equation}\label{up} \nonumber
U^{\pm}(\theta)=
\frac{-2\,K}{2\,\delta\pm\sqrt{-2\,(k+\delta^2)}\,{\rm
cos}[\sqrt{-K}\,\theta]}
\end{equation}
for $c>1$, and
\begin{equation}\label{uh} \nonumber
U^{\pm}(\theta)=\frac{-
2\,K}{2\,\delta\pm\sqrt{-2(k+\delta^2)}\,{\rm
cosh}[\sqrt{K}\,\theta]}
\end{equation}
for $c<1$. Therefore, the solution of Eq.~(\ref{mb}) can be obtained
from (\ref{sgb2}) considering $k=(1-c)/c$, $\theta=x-c\,t$.

We can  also deal with the case $C_{2}\neq0$, $\delta=0$ and the
corresponding forth order polynomial is
\begin{equation}\label{f4}
f(U)=\frac{1}{2}U^{4}+k\,U^{2}+C_{2}
\end{equation}
has four different roots:
\begin{equation}\label{phi04}
U_{0}^{\pm}=\pm\sqrt{-k-\sqrt{-2\,C_{2}+k^2}}
\end{equation}
and
\begin{equation}\label{phi14}
U_{0}^{\pm}=\pm\sqrt{-k+\sqrt{-2\,C_{2}+k^2}}.
\end{equation}
Substituting (\ref{f4}) in terms of $U_0$ in  Eq.~(\ref{x}), the
solution of Eq.~(\ref{ford}) with $\delta=0$ for each root is
\begin{equation}\label{gs4}
U=-\frac{5 \,k \,U_{0}+3\,U_{0}^3+ 12\,U_{0}
\wp(\theta;g_2,g_3)}{k+3\,U_{0}^2-12\,\wp(\theta;g_2,g_3)}
\end{equation}
where
\begin{equation}\label{g4}
g_{2}=\frac{k^2}{12}+\frac{C_{2}}{2},\qquad
g_{3}=-\frac{k^3}{216}+\frac{C_{2}\,k}{12},
\end{equation}
and $\Delta\neq0$.

Now, it is easy to see that  the trivial choice $C_{2}=0$ gives rise
to the invariants (\ref{g2}) with $n=2$
and $\Delta=0$. Then we will get the solitary wave and the periodic
solutions. The roots ($U_0$) also take the forms: $0$ (multiplicity
two) and $U_{0}^{\pm}=\pm\sqrt{-2\,k}$. Therefore, the Weierstrass
function can be expressed in terms of hyperbolic and trigonometric
functions. Then, the solitary wave and trigonometric solutions of
Eq.~(\ref{mb}) can be read from (\ref{sgbs}) and (\ref{sgbp}) with
$n=2$ only for $U_{0}^{\pm}$.

For this case, we have also another type of solution for certain
values of $C_{2}=k^{2}/2$ for which we have the special form of
(\ref{ef11}):
\begin{enumerate}
\item
If $c<1$, the equation
\begin{equation}\label{tan1}
    \left(\frac{dU}{d\theta} \right)^2=\frac{1}{2}(U^{2}+k)^2
\end{equation}
has the solution
\begin{equation}\label{tan11}
U=\sqrt{k}\,\tan\left[\sqrt{\frac{k}{2}}\,\theta\right] .
\end{equation}
\item
If $c>1$, the equation
\begin{equation}\label{tanh1}
    \left(\frac{dU}{d\theta}\right)^2=\frac{1}{2}(U^{2}-(-k))^2
\end{equation}
has the solution
\begin{equation}\label{tanh11}
U=\sqrt{-k}\, \tanh\left[\sqrt{\frac{-k}{2}}\,\theta\right].
\end{equation}
\end{enumerate}
Thus, for $c<1$ the solutions of (\ref{gb}) are periodic singular
kink type
\begin{equation}\label{sgb2k}  \nonumber
u(x,t)= \sqrt{\frac{3(1-c)}{a}}\, \tan\left[
\sqrt{\frac{1-c}{2c}}(x-c\,t)\right]
\end{equation}
and for $c>1$ the solutions are of kink type
\begin{equation}\label{sgb2p}  \nonumber
u(x,t)= \sqrt{\frac{3(c-1)}{a}}\ \tanh\left[
\sqrt{\frac{c-1}{2c}}(x-c\,t)\right] .
\end{equation}

\section{Conclusions}
In this paper we have first factorized the generalized BBM equation
in two ways. Then, we have investigated the travelling wave
solutions of this equation by means of the factorization technique.
We have obtained particular solutions of the generalized BBM as well
as general solutions of the modified-BBM and BBM equations in terms
of elliptic functions without making any ansatz
\cite{wazwaz,wazwaz1,nickel}. We want to stress that this technique
is more systematic than others previously used for the analysis of
these equations. The factorization technique gives directly
solutions of the BBM in terms of elliptic functions. Indeed, we have
more general solutions and recovered all the solutions reported in
\cite{wazwaz,wazwaz1,nickel}. At the same time we have shown the
equivalence of certain expressions that, in fact, describe the same
solutions for the BBM equation.

 \section*{Acknowledgments}
This work has been partially supported by Spanish Ministerio de
Educaci\'on y Ciencia (Projects MTM2005-09183, FIS2005-01375 and
FIS2005-03989), Ministerio de Asuntos Exteriores (AECI grant
0000169684 of \c{S}.K.), and Junta de Castilla y Le\'on (Excellence
Project VA013C05). \c{S}.K. acknowledges Department of Physics,
Ankara University, Turkey, and the warm hospitality at Department of
Theoretical Physics, University of Valladolid, Spain, where this
work has been carried out.
 \section*{Appendix: Elliptic functions}
Let us consider any quartic polynomial
\begin{equation}\label{ef}
f(\varphi) = a_{0}\,\varphi^4+4\,a_{1}\,\varphi^3+6\,a_{2}\,\varphi^2+4\,a_{3}\,\varphi+a_{4}
\end{equation}
and the differential equation
\begin{equation}\label{efd}
\left(\frac{d\varphi}{dt}\right)^2 = f(\varphi).
\end{equation}
The simplest case in which Eq.~(\ref{efd}) is integrable is when $f(\varphi)$ is given by
\begin{equation}\label{ef1}
f(\varphi) =\alpha_{0}^2\,(\alpha^2\pm \varphi^2)^2.
\end{equation}
Then the corresponding differential equation
\begin{equation}\label{ef11}
\left(\frac{d\varphi}{dt}\right)^2=\alpha_{0}^2\,(\alpha^2\pm
\varphi^2)^2
\end{equation}
has the solutions 
\begin{equation}\label{sth}
\varphi^{-}= \alpha \,\tanh[\alpha\, \alpha_{0}\,\theta]
\end{equation}
and
\begin{equation}\label{stan}
\varphi^{+}=\alpha \,\tan[\alpha\, \alpha_{0}\,\theta].
\end{equation}

In the general case, the invariants of (\ref{ef}) are defined as
\begin{eqnarray}
g_{2}&=& a_{0}\,a_{4}-4\,a_{1}\,a_{3}+3\,a_{2}^2,
\\
\label{g3} g_{3}&=&
a_{0}\,a_{2}\,a_{4}+2\,a_{1}\,a_{2}\,a_{3}-a_{2}^{3}-a_{0}\,a_{3}^2-a_{1}^{2}\,a_{4},
\end{eqnarray}
and the variable $z=\int_{\varphi_0}^{\varphi}[f(t)]^{-1/2}dt$, where $\varphi_0$
is any root of the equation $f(\varphi)=0$, is introduced. If the Weierstrass
function $\wp(z;g_{2},g_{3})$ is constructed with the help of the invariants $g_2$ and
$g_3$, then $\varphi$ can be expressed as a rational function of it as
\begin{equation}\label{x}
\varphi=\varphi_0+\frac{1}{4}f'(\varphi_0)\left(\wp(z;g_{2},g_{3})-
\frac{1}{24}f''(\varphi_0)\right)^{-1},
\end{equation}
where the prime ($'$) denotes the derivative with respect to
$\varphi$. We have also the following useful relation for the
Weierstrass function (once the invariants are fixed, and we can
avoid them to alleviate the notation)
\begin{equation}\label{hp}
\wp(z+\omega;g_2,g_3)=e_{1}+\frac{(e_{1}-e_{2})(e_{1}-e_{3})}{\wp(z;g_2,g_3)-e_1}
\end{equation}
where $\omega$ is half-period and $e_{1},\,e_{2},\,e_{3}$ are roots
of the equation $4t^3-g_{2}t-g_{3}=0$, such that
\begin{equation}\label{ee}
\begin{array}{l}
e_{1}+e_{2}+e_{3}=0, \\
e_{1}\,e_{2}+e_{1}\,e_{3}+e_{2}\,e_{3}=-\frac{1}{4}g_{2},  \\
e_{1}\,e_{2}\,e_{3}=\frac{1}{4}g_{3}.
\end{array}
\end{equation}
When $g_{2}$ and $g_{3}$ are real and the discriminant
\begin{equation}\label{di}
\Delta=g_{2}^{3}-27\,g_{3}^{2}
\end{equation}
is positive, negative or zero, we have different behaviors of
$\wp(z)$.

Here, we shall discuss the case $\Delta=0$, that corresponds
to degenerate cases of the Weierstrass functions which occur when
one or both of periods become infinite, or, what is the same, two or
all three roots  $e_{1},\,e_{2},\,e_{3}$ coincide.

If $e_{1}=e_{2}=b>0,\, e_{3}=-2\,b$, then $g_{2}>0$, $g_{3}<0$ and
the real and imaginary periods of the Weierstrass function, $\omega$
and $\omega'$, are $\omega=\infty,\,\omega'=i\pi(12\,b)^{-1/2}$, and
this function can be written as
\begin{equation}\label{sh}
\wp(z;12\,b^2,-8\,b^3)=b+3\,b \sinh^{-2}[(3\,b)^{1/2}z],
\end{equation}
which leads to solitary wave solutions.

If $ e_{1}=2\,b>0$, $e_{2}=e_{3}=-b$, then $g_{2}>0$, $g_{3}>0$,
$\omega=\pi(12\,b)^{-1/2}$, $\omega'=i\infty$, and the Weierstrass
function becomes
\begin{equation}\label{sn}
\wp(z;12\,b^2,8\,b^3)=-b+3\,b \sin^{-2}[(3\,b)^{1/2}z],
\end{equation}
which leads to periodic solutions \cite{Bateman,watson}.

\end{document}